\documentclass[aps,pra,twocolumn,10pt,superscriptaddress,showpacs,longbibliography]{revtex4-1}
\usepackage{hyperref,xcolor,graphicx,enumitem,array}
\usepackage[siunitx]{circuitikz}

\newcommand{\Quote}[1]{\begin{itemize}\item[]#1\end{itemize}}
\usepackage{blindtext}
\usepackage{natbib}
\usepackage{epstopdf}
\usepackage{amsmath}
\usepackage{fullpage}
\usepackage{siunitx}

\begin{document}
	
	% -----------------------------------------------------------------------------------------------------------------------------------
	% TITLE AND ABSTRACT
	% -----------------------------------------------------------------------------------------------------------------------------------
	
	\title{Using think-aloud interviews to characterize model-based reasoning in electronics for a laboratory course assessment.}
	
	\author{Laura R\'ios}
	\email{laura.rios@colorado.edu}
	\affiliation{Department of Physics, University of Colorado Boulder, Boulder, CO 80309, USA}
	\affiliation{JILA, National Institute of Standards and Technology and University of Colorado Boulder, Boulder, CO 80309, USA}
	
	\author{Benjamin Pollard}
	\affiliation{Department of Physics, University of Colorado Boulder, Boulder, CO 80309, USA}
	\affiliation{JILA, National Institute of Standards and Technology and University of Colorado Boulder, Boulder, CO 80309, USA}
	
	\author{Dimitri R. Dounas-Frazer}
	\altaffiliation[Current address: ]{Department of Physics and Astronomy, Western Washington University, Bellingham, WA 98225, USA}
	\affiliation{Department of Physics, University of Colorado Boulder, Boulder, CO 80309, USA}
	\affiliation{JILA, National Institute of Standards and Technology and University of Colorado Boulder, Boulder, CO 80309, USA}
	
	\author{H.\ J.\ Lewandowski}
	%\email{lewandoh@colorado.edu}
	\affiliation{Department of Physics, University of Colorado Boulder, Boulder, CO 80309, USA}
	\affiliation{JILA, National Institute of Standards and Technology and University of Colorado Boulder, Boulder, CO 80309, USA}
	
	\date{\today}
	
	\begin{abstract}
		Models of physical systems are used to explain and predict experimental results and observations. The Modeling Framework for Experimental Physics describes the process by which physicists revise their models to account for the newly acquired observations, or change their apparatus to better represent their models when they encounter discrepancies between actual and expected behavior of a system. While modeling is a nationally recognized learning outcome for undergraduate physics lab courses, no assessments of students' model-based reasoning exist for upper-division labs. As part of a larger effort to create two assessments of students' modeling abilities, we used the Modeling Framework to develop and code think-aloud problem-solving activities centered on investigating an inverting amplifier circuit. This study is the second phase of a multiphase assessment instrument development process. Here, we focus on characterizing the range of modeling pathways students employ while interpreting the output signal of a circuit functioning far outside its recommended operation range. We end by discussing four outcomes of this work: (1) Students engaged in all modeling subtasks, and they spent the most time making measurements, making comparisons, and enacting revisions; (2) Each subtask occurred in close temporal proximity to all over subtasks; (3) Sometimes, students propose causes that do not follow logically from observed discrepancies; (4) Similarly, students often rely on their experiential knowledge and enact revisions that do not follow logically from articulated proposed causes.

	\end{abstract}

	\maketitle
%---------------------------------------
%               INTRODUCTION
%---------------------------------------
\section{Introduction}
\label{sec:Intro}	
% Calls to Action

Recently, the National Research Council (NRC)  called for increased attention to assessments of experimental physics practices, and to the  assessment development process in particular \cite{DBER2012}. One important experimental physics practice is modeling: the construction, testing, use, and revision of models of physical phenomena and apparatus. According to the recommendations released by the American Association of Physics Teachers (AAPT), modeling should be a focus of physics laboratory courses \cite{AAPT2015}. 

Currently, there is no published, validated instrument to assess modeling in physics laboratory courses. There are assessments that measure other concepts in labs, such as ``critical thinking'' \cite{Quinn2017}, how students perceive experimental uncertainty \cite{Buffler2001}, and students' views about experimental physics \cite{Wilcox2016a}. There are also several assessments for laboratory courses in STEM generally (see, e.g., Refs. \cite{Galloway2015} and \cite{Corwin2017} for examples from chemistry and biology). Prompted partly by the NRC recent call to action and the AAPT guidelines for labs, our approach to this work is to build on an assumption that labs are inherently valuable learning spaces, and that there is a need to invest in their improvement \cite{APSbackpage}. The creation of research-based assessments is a critical step in the process to realize this improvement. 

%this is a 4-phase development approach
To address the national calls by NRC and AAPT, our team has begun an extensive process to create two scalable assessments for model-based reasoning in physics lab courses that will be generalizable for use at various levels and institutions. There will be one assessment each for electronics and optics courses. The assessment development process has four phases: 1) determine domain-specific test objectives \cite{DRDF2018}; 2) characterize how students navigate a lab-practicum style activity and justify their choices during think-aloud problem-solving (TAPS) interviews; 3) create a free-response assessment with input from the TAPS interviews and expert physicists; and finally, 4) create a closed-response format assessment based on students responses from Phase 3. The final instrument will have a similar deployment model as the Colorado Learning Attitudes about Science Survey for Experimental Physics (E-CLASS) assessment, and will result in a centralized online survey tool with automated administration \cite{Wilcox2016}. 

\begin{figure*}[t]
	\includegraphics[scale=0.833]{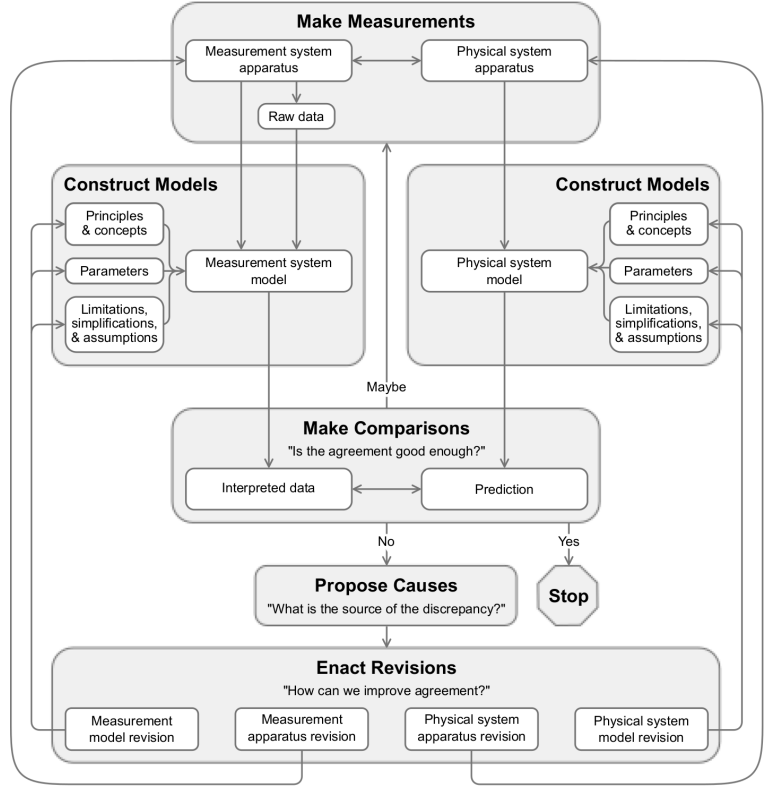}
	\caption{Modeling Framework for Experimental Physics: Originally conceptualized Zwickl \textit{et al.} \cite{Zwickl2014} this schematic of the Modeling Framework from Dounas-Frazer \cite{DRDF2018} shows how five interconnected tasks---Construct Models, Make Measurement, Make Comparisons, Propose Causes, and Enact Revisions---describe the ways scientists bring data and predictions from models into agreement. In the Modeling Framework, we distinguish between physical models/apparatus (e.g., models of the op-amp) and measurement models/apparatus (e.g., models of digital multimeters).}  	
	\label{fig:framework}
\end{figure*}

We are using the Modeling Framework for Experimental Physics to guide all phases of the assessment development. The Framework describes the process by which physicists bring measurements and predictions from models into agreement \cite{Zwickl2015}. The Framework is composed of several interconnected subtasks, depicted in gray boxes in Fig.\ \ref{fig:framework}: making measurements (Make Measurements), constructing models (Construct Models), making comparisons (Make Comparisons) between data and predictions to identify discrepancies, proposing causes (Propose Causes) for those discrepancies, and enacting revisions (Enact Revisions) to resolve them\footnote{We capitalize the names of the modeling subtasks when referring to them as operationalized subtasks. The names appear in lower case when referring to an action performed by a student, e.g., a student makes a comparison between data and prediction.}. 

%this is phase 2: methodology and outcomes
In this paper, we describe the findings from Phase 2 of the assessment development. In Phase 2, we used outcomes from Phase 1 to develop a hands-on electronics activity meant to elicit a large range of modeling behavior. This activity was designed to examine student reasoning around the functional limits of operational amplifiers (op-amps), while allowing for several model and apparatus revisions. We conducted 10 TAPS interviews with upper-division physics and engineering physics undergraduate students at the University of Colorado Boulder (CU Boulder). After coding and analyzing the data, we determined several concrete implications for Phases 3 and 4.

%outline of the paper 	
In Section \ref{sec:research questions}, we outline the three research questions for this paper to describe our overarching goals. In Section \ref{sec:intro-bkg research}, we describe the background research on modeling that we use to broadly contextualize our findings. We also describe specific outcomes from Phase 1 and our team's work on troubleshooting, and how those outcomes have informed activity design and research goals. In Section \ref{sec:methods}, we describe the activity design, interview protocol, and how we coded and analyzed the audio-video data. In Section \ref{sec:analysis}, we present our results of the data analysis according to the research questions. Finally, in Section\ref{sec:discuss} we look forward to Phases 3 and 4 of the assessment development by drawing connections between the outcomes of Phase 2 and concrete ways to incorporate those in subsequent phases. We also contextualize our findings with broader work into students' model-based reasoning, in particular how this work further describes students' difficulty in proposing probable causes for discrepancies. We summarize in Section \ref{sec:conclusion}. 

\section{Research questions}\label{sec:research questions}

The research questions for Phase 2 were developed looking ahead to the structure and content we required for Phase 3. Specifically, we needed to understand how students navigate the modeling process to create relevant test items. We also incorporated instructor feedback from Phase 1 to make sure our final instrument and associated findings aligned with instructor values. 

With that in mind, our research goals were distilled to the following questions about students' approaches to modeling during the TAPS interviews: 

\begin{itemize}
	\item[\textbf{RQ1.}] When students work on an op-amp circuit, how much time do they spend on each of the modeling subtasks?
	\item[\textbf{RQ2.}] When students work on an op-amp circuit, which pairs of subtasks are temporally connected?
	\item[\textbf{RQ3.}] When students work on an op-amp circuit, what types of apparatus and model revisions, comparisons, proposed causes, and measurements do the students make? 
\end{itemize}

We want our final assessment to reflect what students spend time on, but we should not exclude less common subtasks, as they may still play a crucial role in the modeling process. Further, our final assessment instrument will need to have discriminatory power, i.e.\ differentiate between different levels of modeling ability. 

Similarly, we also want to identify common, missing, or unexpected modeling pathways. Accordingly, RQ1 and RQ2 together seek to describe which subtasks students engage in most commonly, and to characterize typical pathways within the context of the Modeling Framework. RQ3 will provide important contextual information for specific scenarios and test items to be developed in Phase 3, the free-response assessment. We refer to these three research questions throughout, and organize the results and discussion sections according to outcomes from each research question.

\section{Select background research}
\label{sec:intro-bkg research}

Both electronics and modeling are active areas of study in physics education research, with a variety of researchers engaged with distinct aspects of both. 

Previous research on students working on circuits tended to focus on conceptual understanding of electronics physics concepts (e.g., Ref \cite{vandeBogart2018}). Recently, Liu and El Turkey analyzed video-taped student performance tests on dc circuits, and concluded that video analysis could help laboratory instructors visualize, and evaluate, students' thinking processes and laboratory skills \cite{Liu2017}. Researchers have also observed how socially mediated metacognition plays a role in troubleshooting malfunctioning electronics circuits \cite{VanDeBogart2017}.

We also draw on previous research into modeling. In the following subsections, we focus on a few specific instances where researchers studied how students engage with measurements and modeling when confronted with a challenging task. Then, we describe the foundational work done by the Lewandowski group at CU Boulder that we drew on explicitly for activity design and goals. Finally, we discuss outcomes from previous studies into troubleshooting electronics circuits, and Phase 1 of the assessment development process.

\subsection{Research on modeling and measurements}\label{subsec:research on modeling}
Models and modeling have been researched extensively in physics education research, and so there are various definitions and conceptualizations of the modeling process, and its role in physics. Here, we describe a few studies whose findings align with and complement our operationalization of modeling.

%importance of measurements
Some previous research in modeling focuses on the role of measurements. Koponen and others have discussed the link between predictions and measurements \cite{Koponen2007}, and have gone further in describing \textit{generative} modeling, wherein measurements are conceptualized as ``investigative'' \cite{Koponen2014}. In this context, investigative measurements can be used to probe concepts, models, and expectations throughout the modeling process, thereby generating knowledge.   

Russ and Odden's work also focuses on measurements. They  describe how students combine evidence-based reasoning and modeling to learn about physical phenomena, and work through initially confusing problems. One of their findings is that students use evidence to further elaborate on model components, and that students' models help guide the search for new evidence \cite{Russ2017}. 

In our work, we combine Russ and Odden's findings on students' use of evidence with Koponen's description of ``investigative'' measurements to situate the role of measurements.  

Another way of conceptualizing an investigative measurement is to understand its outcome. Along this vein, Vonk et al.\ define applying a model to express a quantitative relationship between variables as a part of ``model-making,'' and the necessary revision to the model after critical testing as ``model-breaking,'' both crucial scientific skills \cite{Vonk2017}. Students in that study were primed to consider either model-making, or model-breaking while working on web-based lab activities. For our study, we expect both model-making and model-breaking behavior. An obvious difference is that our work will be conducted with students working on physical apparatus, not an online module, so we may expect more interaction with models of the apparatus itself. 

Similarly, Allie and colleagues have long grappled with students' understanding of the role of measurements, and how to compare two measurements, or sets of measurements, to one another \cite{Allie1998, Buffler2001, lubben2001,Buffler2009}. In one particular study from this group, Volkwyn and colleagues explored what physics majors in traditional introductory physics lab courses think about the nature of a scientific measurement \cite{Volkwyn2008}. One of their findings centered on how and why students came to understand if data sets were in agreement, an area referred to as data comparison. Using the Physics Measurement Questionnaire (PMQ) to determine how students' thinking in this area shifted before and after lab instruction, they found that students' thinking shifted the least toward more favorable reasoning in data comparison when compared to how student reasoning shifted for data collection and data processing. In the Modeling Framework, the Make Comparisons subtask determines the stop criteria, so we are generally interested in ways that these two frameworks coalesce in the realm of data comparison.

\subsection{Research on the Modeling Framework}\label{subsec:research on MF}

Our own research into aligning lab courses with authentic physics experience and practice led to the development and use of the Modeling Framework for Experimental Physics (Fig.\ \ref{fig:framework}) in physics laboratory courses and clinical settings at the University of Colorado Boulder \cite{Zwickl2013,Zwickl2014,Zwickl2015,Stanley2017,Dounas-Frazer2016a,Lewandowski2015}. For a detailed summary of the development and applications of the Modeling Framework, see Ref \cite{DRDF2018Modeling}.

Particularly important for our study is previous work that has identified the Propose Causes subtask as particularly difficult. Zwickl and colleagues found that students' relatively limited theoretical domain knowledge (i.e., physics principles and concepts) hindered their ability to fully explore the limitations and assumptions of their models while working with a photodiode \cite{Zwickl2015}. In a recent publication, R\'ios et al.\ unpacked the varied ways students skip the Propose Causes subtask in the Modeling Framework in favor of other modeling subtasks \cite{Rios2018PERC}. That work aligns with our findings here, and we will show there is a need to understand the circumstances under which a student does or does not propose a probable cause to motivate revisions. 

Further, difficulty in proposing causes can create difficulty in other aspects of modeling, or interrupt the modeling process altogether. Another study on how students document modeling in their lab notebooks found that even when the modeling process is scaffolded by course materials, the students ``generally did not provide [in their lab notebooks] actionable ways of implementing these proposed revisions'' \cite{Stanley2017}. This indicates a difficulty in the Enact Revisions subtask stemming from an unarticulated proposed cause. 

\subsection{Troubleshooting in electronics}

In another related study, students' model-based reasoning was observed while troubleshooting electric circuits \cite{Dounas-Frazer2016a}. Dounas-Frazer et al.\ describe the modeling process during troubleshooting as involving decisions about ``which measurement to perform, in what order, and for what purpose'' (p.\ 5). In their description of troubleshooting episodes, the initial formulation of the problem description is characterized partly by understanding the issues using formative measurements; testing the apparatus using diagnostic measurements to determine the source of a discrepancy; repairing the apparatus; and finally establishing that the overall function of the circuit is correct using evaluative measurements. Our findings here show similar ways of using different types of measurements. 

We also draw on this work to inform our activity design. Dounas-Frazer et al.\ incorporated errors in circuit construction in their activity design, which led to several iterative cycles of figuring out how to fix the circuit. In addition to comparing their predictions to the observed output signal, the students also had to repair the circuit. To explore a different space, we chose to explore measurement/test apparatus errors, where slew rate, bandwidth, and clipping were the main issues, and not circuit construction errors. So, while previously we focused on faulty apparatus that created a malfunction, this work focuses more on aspects of input signals that push op-amps outside of ideal circuit behavior. In this way, we hope to address different modeling pathways by deliberately introducing a distinct context and types of errors. 

\subsection{Phase 1 Outcomes}
Our assessment development process is incremental, with each phase building on the previous. We began by gathering electronics and optics instructor perspectives of modeling in their laboratory courses in Phase 1 \cite{DRDF2018}. The interviews and subsequent analysis yielded how instructors used and valued the modeling process. We also determined that assessments that are aligned with specific contexts (e.g., electronics) need to be developed, since modeling is perceived and used differently in different physics subject areas. For example, electronics instructors often called their courses a ``10\% science'' (in reference to measurement accuracy) or a ``yes-no subject'' (p.\ 14), which informed how students evaluate whether a match to prediction is good enough to consider the discrepancy reconciled.

In the Phase 1 study, we were also interested in common ways the inverting amplifier configuration is used in circuits, and how students model op-amps and op-amp circuits in electronics labs. This was to ensure that our activity can be used by a diverse set of institutions by addressing contexts and models used widely. During the interviews with the electronics instructors, we focused on lab activities they conducted with inverting amplifier circuits. We learned that passive analog electronic circuit elements, such as resistors and capacitors, were commonly used, as were oscilloscopes, digital multimeters (DMMs), and function generators. Instruction on op-amp models were often based on knowing the ``golden rules'' of op-amp operation and their implications. That is, the op-amp may be regarded as a black box in which the inputs have infinite input impedance, and will adjust its output so that the voltage difference between the two inputs is zero while under negative feedback. Op-amps were commonly used to create active filters, amplifiers, and rectifiers.

In terms of the physical and measurement apparatus and phenomenon that were commonly used, an important outcome from Phase 1 was that ``[t]here is no clear evidence for designing an instrument that targets only one particular subtask'' of the Modeling Framework \cite{DRDF2018}---that is, Phase 2 should seek details about how students navigate all the subtasks. Coupled to this finding was the instructors' descriptions of the importance of the Making Comparisons and Make Measurements subtasks, and their observation that students experience difficulties in the Propose Causes subtask. Thus, in Phase 2 we also seek to enumerate and describe the comparisons, measurements, and proposed causes students may undertake. Further, Phase 2 research goals were also informed by the emphasis in Phase 1 ``that the assessment should elicit information about students' justifications for prioritizing some modeling pathways over others'' (p.\ 19).

%---------------------------------------
%               METHODS
%---------------------------------------
\section{Methods}\label{sec:methods}

To collect the type and amount of detail required to understand our three research questions, we gathered data from video-taped TAPS interviews. As discussed in Section \ref{sec:Intro}, a component of the NRC's call for increased attention to assessments for experimental practices is a focus on the development process. In this section, we detail how we incorporated prior research outcomes from Section \ref{sec:intro-bkg research} into the design of the problem-solving activity. We also describe the coding scheme development process, and how our research questions informed the subsequent data analysis. 

\subsection{Activity design}\label{subsec:activity design} 
Common operational amplifier (op-amp) models treat the integrated chips as ``blackbox'' circuit elements \cite{DRDF2018}.  That is, solid-state physical models of the transistors that make up the op-amp are not extensively considered. Practically, it behaves according to gain equations derived from the golden rules, with a frequency dependence that does not follow from the golden rules \cite{Horowitz:1989:AE:76734}. 

%So, we tailored the activity to focus on models of op-amp chip operation and their practical limitations instead of the internal circuitry of the op-amp chip. For example, we did not consider the finite input current of an op-amp. 

\begin{figure}
	\includegraphics[width=1\linewidth]{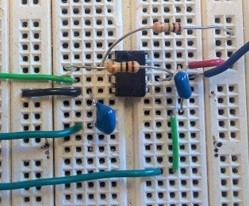}
	\caption{Close-up picture of the circuit as initially constructed by the researchers. The op-amp is an LF356 series. $R_f = 10 \; \si{\kilo\Omega}$, $R_{\text{in}} = 1 \; 1 \si{\kilo\Omega}$ and $C = 100 \; \si{\pico\farad}$.}
	\label{fig:circuit}
\end{figure}

\begin{figure*}%
	\begin{minipage}{0.45\textwidth}
		\begin{circuitikz} \draw
			(0, 0) node[op amp] (opamp) {} %center op amp first, as a node
			(opamp.-) to[R, l_=1<\kilo\ohm>] (-4, 0.5) % using the inverting input of the op amp, use as a coordinate to extend input resistor to the right, hence the negative x value
			(opamp.-) to[short,*-] ++(0,1.5) coordinate (Rf) % feedback resistor
			to[R, l= 10<\kilo\ohm>] (Rf -| opamp.out) %connected to inverting input of op amp
			to[short,-*] (opamp.out)
			(opamp.out)  --++(0.5,0) node[right] {$V_{out}$}
			(opamp.+) --++(0,-0.5) node[ground] {}
			(-4, 0.5) to[sinusoidal voltage source, l_=$V_{in}$] (-4, -1.5) 
			(-4, -1.5) -- (-4,-1) node[ground]{}
			;		
		\end{circuitikz}
		\caption{A circuit diagram of the inverting amplifier circuit used in the TAPS interviews. The students were given a handout with this circuit diagram at the beginning of the activity. The values of the feedback and input resistors were written. Purposefully missing are the values of the power rails (initially set at $\pm 10$ \si{\volt}), and the capacitors from each power rail to ground (100 \si{\pico\farad})}
		\label{fig:setup}
	\end{minipage}	
	\hfill
	\centering
	\begin{minipage}{0.45\textwidth}
		\includegraphics[width=1\linewidth]{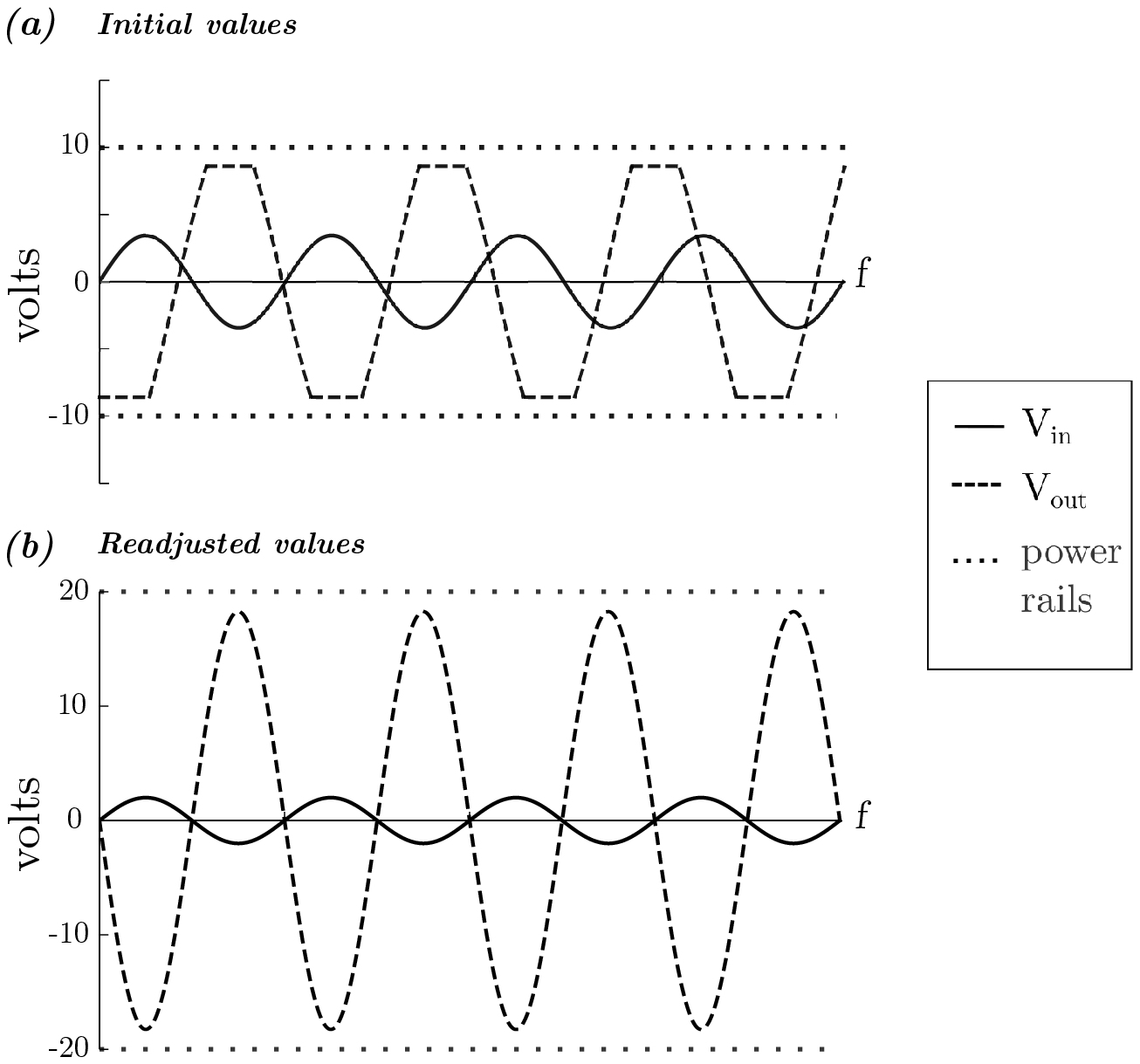}
		\caption{A simulated plot of the input (solid line) and output (dashed line) voltage signals on an oscilloscope with given power rails (dotted line). (a) With the initial conditions, the output signal will appear clipped due to the power rail limit. The slight slant of the output signal is due to slew-rate limits. (b) After hypothetically increasing the power rails and decreasing $V_{\text{in}}$, the clipping and the slight distoratoin due to the slew rate limit is fixed.}
		\label{fig:sawtooth}
	\end{minipage}
\end{figure*}

The central component of our activity was an inverting amplifier circuit (LF356 op-amp) operating outside its recommended voltage and frequency range at a gain of 10 ($A = 10$). We show a picture of the physical circuit as initially constructed in Figure \ref{fig:circuit}, and a circuit diagram is shown in Figure \ref{fig:setup}.

The activity was designed such that there multiple pathways to resolving the issue. Thus, at the onset of the activity, the circuit was powered before the student was instructed to work on it. The frequency ($f$), input voltage ($V_{\text{in}})$, and power rails ($\text{V}_{\text{S}\pm}$) were preset to  $V_{\text{in}} = 3 \; \si{\volt\of{pp}}$, $f = 250 \; \si{\kilo\hertz}$, and $\text{V}_{\text{S}\pm} = \pm 10 \; \si{\volt}$. We included capacitors ($C = 100 \; \si{\pico\farad}$) from the power rails to ground to reduce spontaneous oscillations. Under these conditions, the output signal is clipped (sine wave cut off at the trough and crest) by the power rails, and slightly distorted due to the slew rate limit. Despite these discrepancies, the signal was still recognizable to a trained eye. 

The slew rate limit of the op-amp determines the maximum rate of change of the output voltage, and depends on both the frequency and input voltage amplitude \cite{Horowitz:1989:AE:76734}. Operating above the slew rate limit will result in an output voltage waveform that appears closer to a sawtooth than a sine wave, and with a smaller amplitude. A simulation of the output and input signals at the initial conditions and after readjusting the values is shown in Figure \ref{fig:sawtooth}. 

\subsection{Interview protocol and video data collection}\label{subsec:methods-interview protocol}
At the start of the interview, we provided the student with a schematic diagram of the circuit (Fig.\ \ref{fig:setup}), a data sheet for the op-amp, which included a pinout, and a prebuilt functional circuit. We also included the equation for the gain and reminded the students what to expect for the phase based on the gain equation. 

The interviewer read a short prompt to the student before they began their work on the activity, including a reminder to think aloud while they worked on the circuit. The exact wording can be seen in the interview protocol reproduced in the Appendix.  

The interview included up to two problem-solving activities. The first was to make the circuit function as predicted (correct gain, phase, and shape) with a gain of 10, and the second was to do the same with a gain of 100. The second activity was more difficult to complete than the first. We included the second activity to keep the interview going if the student finished quickly. We also wanted to observe if there were differences in modeling between those students for whom the first activity was straightforward and for whom it was not. Of the 10 participants, five declared they had fixed the circuit, and were given the second activity. Of those that reached the second activity, one completed both activities, and four were still engaged in the second activity at the end of the one-hour interview. Two students did not complete the first activity. Hence, giving the students increasingly difficult tasks after each subsequent hurdle increased the range of the modeling behaviors we observed.

At the end of the TAPS interview, the students were asked a series of questions to gain insight into particular events or modeling tasks that the students might not have articulated aloud, and to understand their academic background in more detail. There were 10 interviews total, all between 55 -- 70 minutes, resulting in about 11 total hours of video data.	

\subsection{Participant recruitment and demographics}
The participants were junior and senior physics and engineering physics majors who had taken the electronics lab in the Physics Department (Electronics for Physical Sciences) during Fall 2016, Spring 2017, or Fall 2017. This class is a standalone, mostly analog electronics laboratory course, as described in Ref \cite{Dounas-Frazer2016a}.

We conducted 10 TAPS interviews during Fall 2017. Students were compensated monetarily for their time. Eight of the ten students answered our final demographic question (``Finally, is it all right if you tell me your gender and your race and/or ethnicity?''). Additional information from the participants was obtained during the post-interview questions. The students' demographic information is presented in Table \ref{tab:demographic}.

\begin{table}[htbp]
	\caption{Demographic Information of participants.}
	\label{tab:demographic}
	\begin{ruledtabular}
		\begin{tabular}{l c }
			Category\footnote{Since the participants were asked to self-identify their gender and race and/or ethnicity, the terms used in this table for those questions reflect the words the participants used, and not categories the research team decided on.}	&  No. of students \\ 
			\hline 
			Engineering physics major  & 1 \\
			Physics major		 & 9 \\
			Third year of study		 & 6 \\
			Fourth year of study		 & 2 \\
			Fifth year of study		 & 2 \\
			Prior research experience & 4 \\
			Male/Men 	 & 6 \\
			Female/Women  & 2 \\
			White/Caucasian & 8 \\
			Did not report race/ethnicity/gender & 2 
		\end{tabular}
	\end{ruledtabular}
\end{table} 

\subsection{Coding scheme and data analysis}\label{subsec:coding} 
We developed an \textit{a priori} coding scheme that was revised during discussions with the research team as a whole, as well as two training cycles of collaborative and independent coding by two authors, L.R.\ and B.P. The cycles of coding improved overall IRR from a Cohen's kappa of 0.69 to 0.88, indicating almost perfect agreement \cite{Landis1977}. After achieving high agreement between the two coders, L.R.\ coded all remaining data and cataloged task-specific emergent subcodes for each code category. We will now describe this process in detail. 

The \textit{a priori} scheme was developed by using the subtasks (gray boxes in Fig.\ \ref{fig:framework}) of the Modeling Framework as the basis for the code categories. L.R.\ created initial drafts of the coding scheme that were then discussed and revised by the research team.

From the audio-video data, the audio was transcribed. The transcripts were synced with the video data. All of the data were coded together in 30-second intervals. The interval size was chosen to capture some of the time-ordering of subtasks, since revisions and measurements in electronics can be done rapidly in succession. 

After the \textit{a priori} coding scheme was created, two authors, L.R.\ and B.P., completed a training phase before the final coding. The training phase consisted of a collaborative coding segment, followed by an independent coding segment, and reconciliation. After the training phase, L.R.\ and B.P.\ separately coded a 10 -- minute segment of the data, then calculated IRR. 

After the first iteration of training and separate coding, we made minor revisions to the wording of the definitions, reconciled differences, and discussed ways in which the robustness of the codebook could be improved. Specifically, we added examples of what to code and what \textit{not} to code. For example, two common revisions were ``Changes the scale on the oscilloscope to get a better look at the signal'', and ``Changes the value of the input voltage, $V_{\text{in}}$ on the function generator''. Changing the scale on an oscilloscope is a technical skill, whereas changing the value of $V_{\text{in}}$ is relevant to the model of circuit behavior. Therefore, we coded changing $V_{\text{in}}$ under the Enact Revisions code, but not changing the oscilloscope scale settings. 

After adding these types of examples, we undertook another round of training and separate coding process, achieving a Cohen's kappa of 0.69. In this round, L.R.\ and B.P.\ noted the code Make Measurements had the most disagreements, in particular on how to incorporate students' use of the oscilloscope to make measurements. In electronics, the oscilloscope is constantly measuring circuit performance, so one might interpret glancing at the oscilloscope as a measurement. However, we wanted to code instances where students were articulating measurements rather than just looking at the oscilloscope. With input from all authors, the code definition was amended to capture the use of utterances about observations made on the oscilloscope, reinforced or contextualized by the video data. 

The following is an example of the revision to the Make Measurements code. In this segment, Sexton is articulating an observation about what she sees on the oscilloscope:

\Quote{ \textit{So now, the gain is only, like 4 again. Yeah, it's not correct.} --Sexton} 

Before revising the codebook, this instance was not coded under the Make Measurements code category. After discussion, we reasoned that Sexton was in fact making a measurement and then a statement about the input and output voltages, determined by eye, on the oscilloscope. All other episodes like the example above were coded under the Make Measurements category. 

\begin{table*}[t!]
	\caption{\label{tab:codes} \textit{A priori} coding scheme and examples from different student participants.}
	\begin{ruledtabular}
		\begin{tabular}{p{2cm} p{8cm}  p{4.5cm}   }
			\multicolumn{1}{l}{Code} & Definition & Example  \\ \hline
			Make measurements        & Student utters or articulates an observation of the data output from a measurement apparatus/device. In the case of electronics, the measurement device may be the DMM, the oscilloscope, or the student's observations about the circuit construction.  
			&  %\textit{Marlowe:} 
			``So the next thing I'm going to do is measure the resistors and make sure they're actually what they say they are.''				\\ \hline 
			Construct Models 		 & Students use their conceptual models of the measurement/physical apparatus/model to reason through the activity. Model construction can occur as a way to orient about the circuit or activity, or after any other subtask has occurred.
			&  %\textit{Sexton:} 
			``They're related because of the gain. Because the gain is Vout over Vin, so, I guess if I'm changing Vin, then the output will change.'' \\ \hline 
			Make comparisons
			& Student compares their observed data to a prediction from the model, or expectations from previous work. The comparison may be qualitative or quantitative. The student may also express that the outcome is ``good enough."
			& %\textit{Lorca:} 
			``They [input and output signals] are definitely between 170 and 180 degrees phase shifted...So I'm going to call that a success.'' \\ \hline 
			Propose causes 			 
			&  Student describes or articulates a reason for the discrepancy between measured/observed quantity and prediction. The proposed cause does not have to have a detailed empirical motivation for it to be coded here.  
			& %\textit{Auden:} 
			``So the problem is that it's operating outside the region, the operation region.''  \\ \hline 
			Enact revisions
			& Student revises, or changes, a component of the measurement/test (e.g., function generator, power supply) or physical apparatus (e.g., resistors in the circuit). The student may also make changes to the measurement or physical model to resolve discrepancies by adding or changing parameters (e.g., may add an arbitrary offset to the gain equation to account for a dc offset). 
			& %\textit{Dickinson:} 
			``What will happen if I change the gain? [...] Right now, I just want to see what'll happen if I switch the resistors.'' \\
		\end{tabular}
	\end{ruledtabular}
\end{table*}

A subsequent round of training and separate coding yielded an overall Cohen's kappa of 0.88, indicating almost perfect agreement. The improvement also indicated that disagreements in the Make Measurements code category were affecting the overall score, and that our revision significantly improved the codebook.  

L.R.\ then completed the rest of the coding using the final coding scheme. An entire 30-second interval would be coded if there was one utterance or sentence that would fit into a code category. Multiple code categories, or no codes, could be assigned to a single 30-second interval. For example, in the following segment, Auden is measuring the resistor values of the feedback and input resistors with a DMM:

\Quote{\textit{This is...1 point 89 (1.89) kilo-ohms. Oops! I mixed up R-f and R-in. This is R-in, that's R-f. Okay, so, that looks pretty okay.} --Auden}

The first portion where he measured the resistor values with a DMM fits under the code category Make Measurements, and his  qualitative statement that the measured values fit his expectation (``that looks pretty okay'') would be considered a comparison. Thus, this 30-second interval would be coded as both the Make Measurements and Make Comparisons code categories. Conversely, if the student is sitting in silence and not performing actions involving the apparatus for the 30-second interval, no codes would be assigned. This was not a common occurrence. The final \textit{a priori} coding scheme, with definitions and examples, is shown in Table \ref{tab:codes}.

During the coding process, we cataloged task-specific, emergent subcodes for all code categories. For example, under the Make Comparisons code category, the act of comparing the expected phase of an inverting amplifying circuit to the observed measurement is a subcode. 

At the end of the coding process, some \textit{a priori} subcodes were deemed too sparsely coded, or too specific to be useful. Therefore, some subcodes were collapsed or combined. For example, the two subcodes for measuring $V_{\text{in}}$ and $V_{\text{out}}$ on the oscilloscope were collapsed into simply, ``Voltage measurement on oscilloscope." The finalized emergent subcodes and their relative frequency in the coding are shown in Table \ref{tab:emergent subcodes}.  

\begin{table*}[!t]
	\centering
	\caption{List of emergent subcodes from the analyzed transcripts. The percentage of coded items refers to how many times a 30-second interval was coded with each code category (subtask). The last column denotes how many interviews out of 10 had at least one instance of the corresponding subcode.}
	\label{tab:emergent subcodes}
	\begin{tabular}{l l  p{2.25cm}  c   }
		Code category            & Subcode                &\% coded items 	& Interviews  \\ \hline
		Construct Models 		 & Relating gain to $V_{\text{in}}$ 
		and $V_{\text{out}}$  & 88               & 9               \\
		& Functional limits 
		of op-amp chip  	      & 12               & 4               \\ 
		\hline
		Make Measurements  		&  Voltage (oscilloscope) & 25               & 10              \\
		&  Shape (oscilloscope)   & 17               & 10              \\
		&  Visual inspection of 
		circuit 			  & 16               & 10              \\
		& Voltage (DMM)			  & 11               & 6               \\
		& Resistor values (DMM)	  & 11               & 8               \\
		& Phase (oscilloscope)	  & 9                & 9               \\
		& ``Noise" (oscilloscope) & 7                & 6               \\
		& Frequency (oscilloscope)& 3                & 6    		   \\ 
		& Current (DMM) 		  & 1				 & 1			   \\
		\hline    
		Make Comparisons    		& Compare circuit 
		construction to pin out   & 22               & 10              \\
		& General goodness of 
		waveform				  & 21               & 10              \\
		& Gain is/is not correct  & 15               & 8		       \\
		& Phase is/is not correct & 11				 & 10 			   \\
		& Sine shape is/is not 
		acceptable 			  & 9				 & 9               \\
		& Student articulates that 
		it is ``good enough."     &  8               & 7    		   \\
		& Predict gain based on 
		values of the resistors   & 4               & 7               \\
		& Predict phase or 
		shape output 			  & 4               & 6     		  \\
		& Predict overall 
		equipment behavior    	  & 3               & 8               \\ 
		& Predict $V_{\text{out}}$ based on 
		gain and $V_{\text{in}}$  & 2               & 5               \\
		& Predict op-amp 
		circuit behavior          & 1                & 2               \\
		\hline  
		Propose Causes			& Values on test/measurement apparatus 
		are too high/low 		  & 34               & 9               \\
		&  Incorrect circuit construction  & 30      & 9       		   \\
		&  Capacitance, resistance, or 
		impedance issue			  & 17               & 8		       \\
		& Test equipment are faulty& 12              & 5               \\
		& Bad op-amp chip 		  & 7                & 4               \\ 
		\hline
		Enact Revisions\footnote{None of the revisions the students enacted were to the model; these revisions correspond only to the measurement or physical apparatus.}
		& Lowers/increases frequency  & 23           & 9                \\
		& Lowers/increases power rails& 22           & 9                \\
		& Lowers/increases  
		$V_{\text{in}}$		          & 20           & 9         	    \\ 
		& Switches out resistors      & 15		     & 7				\\ 
		& Changes function generator
		settings 				       & 7			 & 6				\\ 
		& Re-wires or revises circuitry& 6 		 & 3				\\
		& Changes op-amp chip		  & 5			 & 3       			\\
		
		& Switches to scope probes	  & 1			 & 2		    	\\ 	
		& Changes grounds on power supply & 1		 & 1				\\  
		\hline
	\end{tabular}
\end{table*}

%---------------------------------------
%              RESULTS
%---------------------------------------
	
	\section{Results}\label{sec:analysis}
	Once all of the audio-video data was coded, we analyzed the results. Our analysis was motivated by our three research questions, and so we organize the following subsections accordingly. Specifically, Section \ref{subsec:results-how long} addresses the amount of time spent on each modeling subtask (RQ1); Section \ref{subsec:connections} discusses the findings from examining which subtasks were temporally connected (RQ2); Section \ref{subsec:subthemes} discusses the types of measurements, revisions, proposed causes, and comparisons students made (RQ3). 
	
	\subsection{Time spent on individual subtasks}\label{subsec:results-how long}
	
	The first research question (RQ1) was meant to capture the most prevalent modeling subtasks in electronics. Within this domain-specific assessment, verifying and then interpreting why certain subtasks are more common will aid in developing test items in Phase 3. 
	
	To determine the amount of time students spent on a particular subtask, we aggregated the coded data from all 10 TAPS interviews and counted how many 30-second intervals were coded in the code categories corresponding to the five subtasks in the Modeling Framework: Construct Models, Make Measurements, Make Comparisons, Propose Causes, and Enact Revisions. 
	
\begin{figure}
		\includegraphics[width=1\linewidth]{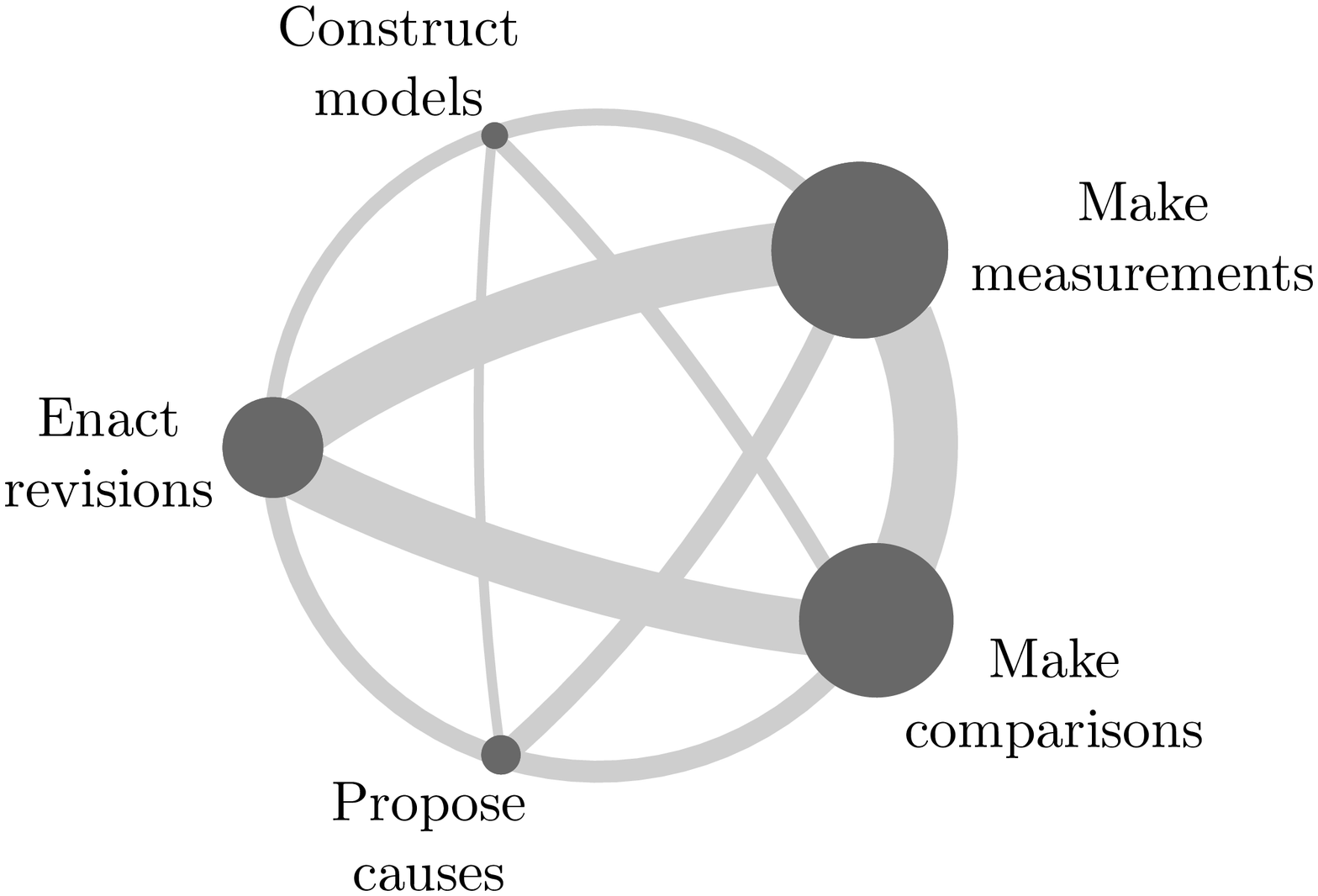}
		\caption{Aggregated data from participants on duration and sequence of modeling subtasks. This schematic shows the commonality of the five modeling subtasks: Construct Models, Make Measurements, Make Comparisons, Propose Causes, and Enact Revisions. The area of the circles represents the number of 30-second intervals coded with that particular subtask; the thickness of the gray connectors between all five subtasks represents the number of times a subtask was coded directly after the other or within the same 30-second interval. This representation is similar to the transition graphs presented in Ref \cite{Cancula2015}.}
		\label{fig:pairwise connections}
	\end{figure}

	In Fig. \ref{fig:pairwise connections}, the aggregated data are shown in a schematic highlighting the commonality of codes and their connections. The area of each dark gray circle represents the number of 30-second intervals that were coded with a particular subtask. The thickness of the gray connectors indicates how many times a subtask was coded directly after another subtask per 30-second interval, in addition to the number of times each pair was coded in the same 30-second interval.  
	
	We found that students spent the most amount of time in the Make Measurements, Make Comparisons, and Enact Revisions subtasks. The large area of the Make Measurements subtask may be partly due to the ease with which one may carry out measurements in electronics. Further, we operationalized the Make Measurements subtask to include utterances about what a student observes on the oscilloscope, which is constantly taking data. However, these potential reasons why Make Measurements is the most commonly coded item are not artificial; this speaks primarily to how measurements are used and interpreted in electronics. How we will capture the significant role of measurements in the assessment will be discussed in Section \ref{subsec:implications for assessment}
	
	Compared to measurements, comparisons, and enacting revisions, students spent much less time proposing causes and constructing models. The lack of items coded in the Construct Models subtask may be partially explained by how much information we intentionally gave the students about fundamental aspects of the op-amp operation model. The students may not have had much opportunity to construct further models, a tension that will be discussed within the context of subsequent phases of the assessment in Section \ref{subsec:implications for assessment}.   
	
	\subsection{Temporal connection of subtasks}\label{subsec:connections}
	
	The second research question, RQ2, seeks to understand which subtasks were clustered together in time to identify and describe common, missing, or unexpected pathways. To this end, we counted the number of times a particular subtask in a 30-second interval (e.g., Make Measurements) was directly followed by another modeling subtask (e.g., Make Comparisons). We also counted the number of times the subtasks were coded within the same 30-second interval (e.g., Make Measurements and Make Comparisons). This includes utterances that would be defined as both subtasks and subtasks that occur rapidly after one another. The gray connecting lines show the connection between just two subtasks, which we have termed \textit{adjacent pairs}. The thickness of the connectors denotes how many times an adjacent pair was observed.  
	
	From the data, we observe that all adjacent pairs are present. We interpret this as representative of the interconnectivity of the modeling process. While the Modeling Framework describes a recursive, nonlinear process, these data also show that students may undertake the modeling subtasks in any order.
	
	Make Measurements $ \leftrightarrow $ Make Comparisons, Make Measurements $\leftrightarrow$ Enact Revisions, and Make Comparisons $\leftrightarrow$ Enact Revisions are the most prevalent adjacent pairs. This, together with the finding in Section \ref{subsec:results-how long} that these three subtasks are also the three most common subtasks in this data set, underscores the centrality of iterative measurements and revisions in electronics. When confronted with a challenging issues, students may skip the Propose Causes subtask in favor of enacting more revisions or taking further measurements (see Ref {\cite{Rios2018PERC}}).  
	
	\subsection{Description of subcodes and themes}\label{subsec:subthemes}
	
	For our third research question (RQ3), we were interested in cataloging which types of measurements, comparisons, proposed causes, and revisions the participants made during the TAPS interviews.
	
	As described in Section \ref{subsec:coding}, task-specific subcodes emerged during the coding process, and were either collapsed or reallocated to better describe the data. We also describe particular motivations or circumstances we noticed that accompanied several of the subcodes. The following sections detail the emergent subcodes and some nuances we observed. These descriptive data will aid in developing relevant test items for Phase 3, and their implications for subsequent phases of the assessment development will be revisited in Section \ref{subsec:implications for assessment}. 
	
	\subsubsection{Make Measurements}
	
	Students constantly make many measurements of different types. There were three general types of measurements we cataloged -- those accomplished with the oscilloscope, the digital multimeter (DMM), and visual inspection. 
	
	Measurements made using the oscilloscope and the DMM constituted the majority of the measurements. Students showed ease with both making the measurements, and interpreting the data. For example, the most common measurements where those verifying the voltage, phase, and shape of the input and output signals by visual inspection on the oscilloscope. Sometimes, students made measurements on the general goodness of the output signal using qualitative descriptors like ``noise,'' ``stable,'' and ``shaky.'' For example, in the following quote, Lorca\footnote{All names are pseudonyms.} made successive measurements about the so-called shakiness of the output signal at different input voltages. 
	
	\Quote{\textit{So now, we got 1 volt input [Vin]. So...[pause] Getting some kind of shakiness at 1 volt from the wave amplifier [high frequency noise pick-up], so I'm going to change it [Vin] to 2 volts. And that seems even more stable.} --Lorca} 
	
	In the following example, Plath is beginning the activity and has just verified that the circuit construction is correct. She then makes additional measurements on the oscilloscope to measure the vlues of $V_{\text{in}}$ and $V_{\text{out}}$.
	
	\Quote{\textit{So, it looks like the...so looks like my peak to peak output is like 5 volts for the input versus...1, 2, 3...I think 3 and a halfish [voltage divisions on the oscilloscope], uh, yeah. So 15, 17.5 volts? So, oh in any case, that's like 3 and a half...a gain of like 3 and a half, which is not 10.} --Plath}
	
	Students used the DMM mostly to make measurements on the circuit--specifically, if the power supply was functioning correctly and powering portions of the circuit. Most students would also check the resistor values with the DMM. 
	
	All 10 participants initiated the activity by visually inspecting their circuit to ensure that it was constructed correctly. Sometimes when they were completely out of ideas, they would revisit the circuit construction as a possible source of error, e.g., 
	
	\Quote{\textit{Okay, I'm going to check that everything's plugged in like it's supposed to be, just to double-check. Input is going from the function generator through the resistor into...positive input is grounded...and, from the negative one fed back into output...} --Frost} 
	
	Of the subcodes encountered in the Make Measurement subtask, students struggled with very few of them, indicating an overall adeptness at making measurements. Engaging with \textit{which} measurements are common and conducted with ease, not just looking at the Make Measurement subtask overall, helps establish guidelines for subsequent phases of the assessment in Section \ref{subsec:implications for assessment}.    
	
	%	[condense the subtlety here to just the baseline vs non-baseline measurements]
	
	\subsubsection{Construct Models}
	
	Overall, there was not much opportunity for model construction, since the majority of an op-amp model was purposefully given to the students, including the equation $A = - R_f/R_{in}$. We did not ascribe students calculating gain from the ratio of the resistors to model construction; this would be more accurately described as \textit{using} a model by parameterizing the gain with the given resistor values. Instead, we used two emergent subcodes to describe the models students did construct. 
	
	One subcode describes constructing models about the output voltage signal, given the gain and the input voltage, using the equation $A = - R_f/R_{in}$. 
	
	Another subcode encapsulates how students will sometimes construct further models on the relationship between $V_{\text{out}}$ and $V_{\text{in}}$ to $A$ with the equation $ V_{\text{out}} = A V_{\text{in}}$. This model construction process is exemplified by the following quote, where Sexton reasons about how to relate $V_{\text{in}}$ and $V_{\text{out}}$ directly based on what she measures on the oscilloscope: 
	
	\Quote{\textit{Actually, yeah, that makes sense. They're related because of the gain. [...] Because the gain is V-out over V-in, so, I guess if I'm changing V-in, then the output will change.} --Sexton} 
	
	The second subcode describes how students construct models of op-amp function when considering the operational limits of op-amp behavior. The model construction around the functional limits of the op-amp included reasoning about frequency, input voltage, and power limitations. Utterances coded here were typically done at the beginning of the activity, likely to assess the initial parameters. In the following example, Lorca constructs a model aloud about how the functional limits of the op-amp will affect the signal he observes:
	
	\Quote{\textit{I guess I understand that whenever something's saturated, it's because you're operating outside the limits of the op-amp.} --Lorca}
	
	Here, the ``saturation'' Lorca mentions refers to the clipping of the output signal due to the limited power rails. He is connecting this observation with a practical model of op-amp circuit behavior by qualitatively identifying that op-amps have functional limits.

	\subsubsection{Make Comparisons}
	
	The Make Comparisons subtask was about as common as Make Measurements. The most common type of comparison was checking the circuit construction to the given pinout, datasheet, and circuit diagram. All participants engaged in these comparisons.
	
	We found that the majority of the comparisons were evaluating the goodness of the gain, phase, and shape, aligned with the goals of the activity as articulated to them at the beginning of the interview. In addition, students would compare the general acceptability of the output waveform to an expectation of noise. All of the students that were able to finish the first prompt articulated when they thought all three output signal characteristics they were asked to fix were ``good enough'' at the end of the activity. 
	
	Within the Make Comparisons subtasks, we noticed that comparisons could be either be qualitative or quantitative. Comparisons of shape, phase, and general goodness would often be qualitative. For example, a student may compare the waveform's qualitative characteristics to some standard of general goodness and shape:
	
	\Quote{\textit{So it's [the output signal] sort of sinusoidal, but mostly it just appears to be noise.} --Frost}
	
	In contrast, students often made quantitative comparisons about gain, or constituent voltages, as in the following example from the same student: 
	
	\Quote{\textit{Right now the input is at 1 volt peak to peak, and then the output is at 10 volts peak to peak, which is a gain of 10, which is what I want.} --Frost} 
	
	In addition to comparing their data to a model in some way, we also noticed that students would articulate comparisons between successive measurements as getting them closer or further from their goal. They may iteratively compare their measurements as they revise possible parameters. In the following example, Dickinson qualitatively compares the ``stability" of their waveform after enacting a revision to the physical apparatus:
	
	\Quote{\textit{Huh. That actually stabilized the waveform a little bit more. [...] I switched the 1 mega-ohm [resistor] with a 1 kilo-ohm resistor, switched the R-f [feedback resistor], made the gain smaller.} --Dickinson}
	
	Embedded in the operationalization of the Make Comparison subtask were the predictions students made. The most common predictions were based on the expected gain from the given values of the resistors, often completed at the beginning of the activity, and at the initiation of the second prompt if the student finished early. Phase and shape predictions were also common. Many predictions were made about the output voltage given the gain and the input voltage from the function generator. Further, model construction around the $V_{\text{out}}$ to $V_{\text{in}}$ ratio often led to comparisons using students' predictions about the output voltage amplitude ($V_{\text{out}}$ based on gain and $V_{\text{in}}$), and the gain calculated from the resistors.
	
	Interestingly, students would some times also predict that the circuit should be completely functional, but it evidently was not (subcode ``Predict overall equipment behavior''). These subcodes are characterized by utterances such as:
	
	\Quote{\textit{It [the circuit] should be able to produce the gain predicted by the resistances. I shouldn't have to change the...predicted gain by changing the resistors.} --Plath}
	
	Here, Plath has just begun the activity and is beginning to reason about the parameters of the gain. She predicts that the absolute resistance values are neither too high or too low, and therefore the gain produced by those resistors is an acceptable value. These predictions played an important role in aiding the student to determine useful measurements or revisions to make if their prediction turned out to be incorrect.

	\subsubsection{Propose Causes}
	
	The types of proposed causes that the students articulated did not vary significantly between different interviews. Most of the proposed causes were about incorrect values on the measurement or test apparatus, e.g., input voltage that was too high for correct operation of the circuit. An equally common proposed cause was incorrect circuit construction, but that was not a designed flaw of the activity. Other typical proposed causes had to do with faulty equipment, issues with impedance or resistance (e.g., the internal resistance of the test apparatus), or a bad op-amp. 
	
	We noticed that many instances of students proposing causes did not exactly address the measurement or comparison the student had just made. That is, a proposed causes many be \textit{connected} or \textit{disconnected}, in reference to whether the uttered proposed cause was connected to the results of the comparison or measurement. We found that often, students would articulate that they understood that \emph{something} was wrong, but did not connect their observations to a specific proposed cause. Indeed, there were several times when students were attempting to understand the nature of a particular discrepancy, and instead of following through with a probable proposed cause, instead seemed to assign a disconnected cause from a list of typical causes, presumably from experiential knowledge. For example, in the following episode, Frost has maneuvered the circuit into a regime where spontaneous oscillations are obscuring the output signal. When he measures the voltage from the output pin on the op-amp to ground with the DMM, the output signal clears up.
	
	\Quote{\textit{So, if I measure from here to here, then it [output waveform] clears up. [measures again, pauses] That is strange. So, I'm trying to figure out what the multimeter does when it's in voltage mode. It is in parallel, and it has a very high resistance. [brief silence] Huh. I'm going to try another op-amp, because I don't...I think it's a different issue.} --Frost}
	
	Frost began to reason correctly about how his measurement device, the DMM, would affect his measurement. Instead of proposing a cause having to do with the actual effect of the DMM ( resistance of the DMM is in parallel with the circuitboard when measuring), he instead opted to propose that there is a ``different issue," which would be addressed by changing the op-amp. These episodes exemplify how a \textit{disconnected} proposed cause could be used to inform a revision. 
	
	Overall, we found that the Propose Cause subtask was the area where students appeared to resort to other subtasks when they were unable to initially propose a cause for an observed discrepancy. Most proposed causes were disconnected, highlighting the difficulty around this subtask. We discuss this further in Section \ref{subsec:implications for assessment}. 
	
	\subsubsection{Enact Revisions}
	Revisions to any part of the experimental apparatus were the most readily undertaken. Accordingly, the most common revisions were to the test/measurement apparatus: the function generator and the power supply. The power rails were revised frequently because, often, the students did not have a complete model of how these voltage levels would affect the signal, and so they would explore the effect of changing these values for a significant amount of time. Changes to the input voltage were similarly exploratory and continuously varying. Instead of trying discrete input voltage values with a proposed effect in mind, they would often search the entire range of voltages available. 
	
	Less commonly, the students would revise extraneous settings on the function generator, such as the dc offset and output impedance settings. Only a couple students revised how the measurement/test apparatus was used, such as using oscilloscope probes to change how they took a voltage measurement. 
	
	Revisions to the physical apparatus were dominated by switching out the resistors. Most of these instances were times when the students reasoned that the initial given values were inappropriate. For example:
	
	\Quote{\textit{I should probably use a different ratio of resistors so that um...Maybe smaller? A smaller value, so that both of them are smaller in order to get more current in? I don't know. So, before I had 10 kilo-ohms to 1 kilo-ohms, so maybe...1 kilo-ohm to 100 kilo-ohms.} --Plath}
	
	The students also revised their physical apparatus by rewiring or revising the circuit construction, or by changing the op-amp chip. Several students changed the op-amp chip more than once. 
	
	Interestingly, we found no evidence that revisions to either the physical or measurement model were made, even though in principle students could have used a revised equation for the gain which includes a frequency dependence.
	
	Similar to the Propose Causes subtask, we noticed that the students often enacted revisions that did not stem directly from a proposed cause, and less often, made revisions that directly addressed either a comparison, or a proposed cause. We also called these \textit{disconnected} or \textit{connected}, respectively. The differences between the motivations or circumstances behind an enacted revision indicates that the difficulty around enacting revisions is related to general difficulties in proposing causes.

%---------------------------------------
%               DISCUSSION
%---------------------------------------
	
	\section{Discussion}\label{sec:discuss}
	{To organize the discussion, we will first synthesize the findings from Phases 1 and 2 to contextualize this paper's findings within the incremental assessment development process, and discuss how we propose to incorporate these findings into the subsequent phases of the assessment in Section \ref{subsec:implications for assessment}. In Section \ref{subsec:limitations}, we elaborate on the limitations of this study, and how we hope to ameliorate some of them in upcoming phases of the assessment development. We close by discussing how our work is synergistic with other work on students' model-based reasoning generally in Section \ref{subsec:future research}).  
		
		\subsection{Implications for design and structure of subsequent assessment phases}\label{subsec:implications for assessment}
		
		%making measurements and comparisons 
		In Phase 1, we learned that instructors found the Make Measurements subtask to be one of the most frequently undertaken by students and an important subtask with which to have practical facility \cite{DRDF2018}. Many spoke about how important it is for students to learn how to use equipment to make measurements (p.\ 12). 	
		
		In our data, we found that the Make Measurements subtask is the modeling subtask that the students undertake most. Additionally, the Make Measurements subtask forms an adjacent pair with every other subtask. When we analyze the subcodes for the Make Measurement subtask, we noticed that there was a subset of measurements with which all students made, mostly around using the oscilloscope to evaluate voltage, phase, and waveform measurements. We found that the strong connections between Make Measurements and other subtasks are partly due to the fact that there are varied reasons why and how a student would undertake a measurement. This observation indicates that measurements dominate the modeling pathway not necessarily just because they are easy to make in electronics, but also because students often need measurements to affirm or discard their conclusions, or motivate why a conclusion is posited. 
		
		In short, the Make Measurements subtask is one that students undertake often, and with ease, and so it is not crucial that we extensively assess student competency with this subtask. However, it is crucial for students to have the opportunity to see the results of measurements to guide their modeling process. Therefore, Phases 1 and 2 together demonstrate a need for an assessment that allows students to easily access the results of measurements that they make often.
		
		For Phase 3 of the assessment, one way to allow students to make measurements is to include an image of an oscilloscope screen that the students can use to measure the output and input signals. It will also be necessary to include the ability to see the result of a measurement with a DMM. The decision of when and what to measure will be up to the student, but after that decision is made, we will provide the result of the measurement. This is in contrast to allowing students to interact with a virtual measurement instrument like in a PhET simulation\cite{mckagan2008developing}. 
		
		%proposed causes 
		Another outcome from our data is that students often struggle to propose causes; it is the modeling subtask they spend the least time on, and consequently, the least likely to form an adjacent pair with other subtasks. This is not altogether unexpected. In Phase 1, both electronics and optics instructors highlighted the fact that their own students seemed to struggle the most with proposing causes for discrepancies. Examining Table \ref{tab:emergent subcodes}, we see that a couple of the most common types of proposed causes---``Incorrect circuit construction,'' and ``Capacitance, resistance or impedance issue''---are not connected to a discrepancy. Indeed, the circuit was constructed correctly, and many instances of the latter proposed cause were vague or speculative. Thus, it seems that students are selecting from a limited range of concepts for both enacting revisions and proposing causes, and that different students have the same limited set of tools. Both research outcomes---short duration of the Propose Causes subtask and disconnected reasoning---highlight the need for assessing the Propose Causes subtask repeatedly, precisely because this subtask appears to be quite difficult. 
		
		%enacting revisions 
		We found that the Enact Revisions subtask was a similarly challenging subtask for students.  A majority of instructors from Phase 1 identified the Enact Revisions subtask as one students undertake rarely, citing mostly time constraints and lack of facility with apparatus. Most instructors described instances of students enacting revisions as part of the troubleshooting process \cite{DRDF2018}. We have previously shown that instructors think troubleshooting is extremely important \cite{DRDF2017troubleshooting} and that a major goal of electronics is for students to build circuits that work \cite{Lewandowski2014}. In addition, instructors believe that since ``nothing works the first time'' \cite{DRDF2016PERC} troubleshooting and revisions are a required part of working with electronics.
		
		In the data from Phase 2, we found that students engage frequently with the Enact Revisions subtask also in the context of troubleshooting. The data from Phase 2  suggest that we should be deliberate about distinguishing between connected and disconnected revisions, as sometimes students will revise the apparatus as a way to understand the discrepancy before proposing a cause. Therefore, Phases 1 and 2 together show that revisions should play a significant role in the assessment and attention should be paid to the subtask's connectedness to a comparison or measurement.  
		
		Looking ahead to Phase 4, the final assessment should also take into account student reasoning for proposing causes and enacting revisions. We had previously proposed to do this by using a coupled-multiple response assessment for Phase 4, where a student is asked to choose a reasoning for their selection in a multiple-choice question (see, e.g., Ref \cite{Wilcox2015}). The results from this study suggest that the available reasoning elements should include an option along the lines of, ``I didn't know what else to do," in order to identify motivations for particular revisions and causes. 
		
		Overall, the outcomes of our data analysis from Phase 2 complement Phase 1 outcomes to create a clearer picture for subsequent phases of the assessment development. As discussed, future phases should probe all five modeling subtasks. From our analysis of the modeling subtask adjacent pairs, we identified the need for a process-oriented assessment that allows students to undertake each subtask in any order. Thus, we have identified the need for a two-part assessment, with one portion focused on the processes and pathways students take, and the one focused on student competency with each modeling subtask. An assessment with this structure will help us probe students' abilities with respect to individual subtasks, and connections between subtasks. 
	}	  
	
	\subsection{Limitations of Phase 2}\label{subsec:limitations}
	
	Even though we were seeking to show the variety of student behavior around modeling an electronics circuit, we found no evidence that students revised their models of either the measurement or physical systems under investigation. We note that students' ability to construct models is closely dependent on their domain knowledge. This suggests that the final instrument necessarily will need to assess domain knowledge to the degree that it allows students to engage in the modeling process. Further, we cannot make any claims about students' revisions of models. This is a significant limitation of this study that we will address by seeking expert physicists' feedback specifically about model revisions in Phases 3 and 4 of the assessment development process. 
	
	In Phase 1 of the assessment, our research team was purposeful about seeking out diverse types of institutions from which to draw instructor interviews. In our vision, the ultimate assessment should be generalizable to various institutional contexts and student populations. 
	
	Due to time and cost constraints, we were not as broad here in recruitment for participants. Specifically, all the student participants for this study were currently enrolled at CU Boulder. CU Boulder is a large, selective R1 (highest research activity), predominantly white institution with a very large physics department \cite{Carnegie2015}. The institutional context introduces limitations to the study, as its results may not translate completely to other student populations and contexts. In contrast, Phases 3 and 4 will involve free-response or coupled-multiple response survey data, which are much more conducive to multi-institution studies. 	
	
	\subsection{Connections to other research directions}\label{subsec:future research}
	
	While the focus of this research paper was to describe the progress towards an assessment, and pay careful attention to its development and process as suggested by the NRC \cite{DBER2012}, we certainly foresee that our findings may inform future research on student understanding of electronics. 
	
	In our study, we described the centrality of measurements in the modeling process. Russ and Odden have also described the role of collecting evidence to make sense of a system \cite{Russ2017}, which adds to our description of measurements as a sensemaking, investigative tool in modeling. We observe similar patterns of student engagement with modeling when we consider measurements as evidence. 
	
	We also observed students making continuous measurements and revisions when confronted with a confusing problem for which they could not propose a cause for discrepancy. These episodes varied in time, but often took several minutes during which the student explored the functional limits of the op-amp by making different types of measurements. In this way, this work is distinct from Vonk et al.\ \cite{Vonk2017} in that the longer timescale of iterative measurements observed in our data set could illustrate extended instances of model-making when students tested the functional limits of the op-amp model. What Vonk calls model-making, we may describe using the language of the Modeling Framework as model construction. However, we did not observe model-breaking, which we would describe as revisions to the model after critical testing. Instead, revisions were made exclusively to the apparatus. However, Vonk's work suggests model-making and model-breaking (loosely construed, model construction and model revision respectively) are necessarily interlocked. In the future, investigating the role of domain knowledge in prompting either model-breaking or revisions to the model could shed light on why we did not observe model revisions.  
	
	Finally, here we will continue to explore the complexity of students' general difficulty with proposing a probable cause for discrepancy. While left unanalyzed in this work, we note the unexpected existence of Propose Causes $\leftrightarrow$ Construct Models adjacent pair. It is interesting to note that some times these two subtasks came together, presumably as a sensemaking tool. Examining these episodes would be a valuable way to understand why and how the Propose Causes subtask is challenging. 
	
	%-----------------------------------------------------------------------------------------------------------------------------------
	% Conclusion
	%-----------------------------------------------------------------------------------------------------------------------------------
	
	\section{Conclusion}\label{sec:conclusion}
	Our team is currently developing two modeling assessments for upper-division physics electronics and optics laboratory courses in four incremental phases. This work describes the goals and outcomes of Phase 2 for the assessment on electronics, and ways our findings productively align with, extend, and complement other work on model-based reasoning. We conducted 10 think-aloud problem-solving interviews with CU Boulder upper-division physics students in order to learn how students engage in model-based reasoning while working on a practicum-style electronics activity consisting of an inverting amplifier circuit. Another goal of this work is to document in detail the assessment development process, keeping in mind the calls of the NRC to create evidence-based assessments of experimental physics practices.   
	
	Four outcomes of this work are: (1) Students engaged in all modeling subtasks, and they spent the most time measuring, comparing, and revising; (2) Each subtask occurred in close temporal proximity to all over subtasks; (3) Sometimes, students propose causes that do not follow logically from observed discrepancies; (4) Similarly, students often rely on their experiential knowledge to enact revisions that do not follow logically from articulated proposed causes.   
	
	To accommodate these outcomes in Phase 3, we propose a two-part assessment with a process-oriented section, and a free response survey to assess competency within individual modeling subtasks. We also propose providing the student with the results of measurements and revisions that we observed here to be easily undertaken, and thus, not relevant to assess.

%-------------------------------------------------------------------------------------------------%----------------------------------
% ACKNOWLEDGMENTS
% -----------------------------------------------------------------------------------------------------------------------------------
\begin{acknowledgments}
	We are grateful to the CU Boulder physics students who volunteered their time to participate in this study. We would like to thank Dr.\ Kevin L.\ Van De Bogart for his suggestions on developing the coding scheme, and Dr.\ Bethany R.\ Wilcox for her suggestions on improving the manuscript. This material is based on work supported by the National Science Foundation under Grant Nos. DUE-1611868, DUE-1726045, and PHY-1734006.
\end{acknowledgments}

% BIBLIOGRAPHY
% -----------------------------------------------------------------------------------------------------------------------------------
\bibliographystyle{apsrev4-1} 

\end{document}